%% file: pub_d0sqgl.tex
\documentclass[aps,prl,twocolumn,showpacs,groupedaddress]{revtex4}  
\usepackage{graphicx}  
\usepackage{dcolumn}   
\usepackage{bm}        
\usepackage{amssymb}   

\newcommand{\ttb}{$t\bar{t}$}
\newcommand{\bbb}{$b\bar{b}$}

\newcommand{\invpb}{pb$^{-1}$}

\newcommand{\met}{\mbox{\ensuremath{\slash\kern-.7emE_{T}}}}

\newcommand{\mht}{\mbox{\ensuremath{\slash\kern-.7emH_{T}}}}

\def\lsim{\mathrel{\rlap{\lower4pt\hbox{$\sim$}}
    \raise1pt\hbox{$<$}}}                

\newcommand{\HT}{H_T}

\newcommand{\etadet}{\vert\eta_{\mathrm{det}}\vert}

\newcommand{\W}{W}
\newcommand{\Z}{Z}

\newcommand{\q}{q}
\newcommand{\qb}{\bar{q}}

\newcommand{\sq}{\tilde{q}}
\newcommand{\sg}{\tilde{g}}

\newcommand{\xo}{\tilde{{\chi}}^0_1}
\newcommand{\xpm}{\tilde{{\chi}}^\pm}

\begin{document}

\hspace{5.2in} \mbox{Fermilab-Pub-06/06-077-E}

\title{Search for squarks and gluinos in events with jets and missing transverse energy\\
in $\bm{p\bar{p}}$ collisions at $\bm{\sqrt{s}=}$1.96~TeV}

\input list_of_authors_r2.tex  
\date{May 14, 2006}

\begin{abstract}
The results of a search for squarks and gluinos using data from $p\bar{p}$ collisions 
recorded at a center-of-mass energy of 1.96~TeV by the D\O\ detector at the Fermilab 
Tevatron Collider are reported. The topologies analyzed consist of acoplanar-jet and multijet events 
with large missing transverse energy. 
No evidence for the production of squarks or gluinos was found in a data sample of 
310~$\mathrm{pb}^{-1}$. 
Lower limits of 325 and 241~GeV were derived at the 95\% C.L. on the squark and gluino 
masses, respectively, within the framework of minimal supergravity 
with $\tan\beta = 3$, $A_0 = 0$, and $\mu < 0$. 
\end{abstract}

\maketitle 

Supersymmetric models predict the existence of spin-$0$ quarks, or squarks ($\sq$),
and spin-$1/2$ gluons, or gluinos ($\sg$), as partners of the ordinary quarks and gluons. 
Supersymmetric particles carry a value of $-1$ for $R$-parity, a multiplicative quantum number,
while $R=1$ for standard model (SM) particles.
If $R$-parity is conserved, as assumed in the following, supersymmetric particles are produced in pairs. 
Their decay leads to SM particles and to the lightest supersymmetric
particle (LSP), which is stable. 
In supersymmetric models inspired by supergravity~\cite{msugra}, the commonly accepted LSP 
candidate is the lightest neutralino ($\xo$, a mixture of the superpartners of the neutral gauge 
and Higgs bosons), which is weakly interacting, thus escaping detection and providing 
the classic missing transverse energy ($\met$) signature at colliders. 
The most copiously produced supersymmetric particles in $p\bar{p}$ collisions should 
be, if sufficiently light, colored particles, {\sl i.e.} squarks and gluinos. 
If squarks are lighter than gluinos, 
they will tend to decay according to $\sq\to\q\xo$, and their pair production will yield 
an acoplanar-jet topology with $\met$. If gluinos are lighter than squarks, their pair 
production and decay {\sl via} $\sg\to\q\qb\xo$ will lead to topologies containing a large 
number of jets and $\met$. 

In this Letter, a search for squarks and gluinos in topologies with jets and large 
$\met$ is reported, using 310\,\invpb\ of data collected at a center-of-mass energy 
of 1.96\,TeV with the D\O\ detector during Run~II of the Fermilab Tevatron $p\bar{p}$ Collider.
The search was conducted within the framework of the minimal supergravity model
(mSUGRA)~\cite{msugra}. 
Previous direct mass limits are 195\,GeV for gluinos if squarks are very 
heavy, and 300\,GeV for squarks and gluinos of equal masses~\cite{Abbott:1999xc,Affolder:2001tc}. 

A detailed description of the D\O\ detector can be found in 
Ref.~\cite{Abazov:2005pn}. The central tracking system consists of a
silicon microstrip tracker and a central fiber tracker,
both located within a 2~T superconducting solenoidal magnet. A liquid-argon 
and uranium calorimeter covers pseudorapidities up to $|\eta|$ $\approx 4.2$, where 
$\eta=-\ln \left[ \tan \left( \theta/2 \right) \right]$ and $\theta$ is the polar angle 
with respect to the proton beam direction. The calorimeter consists of three sections, 
housed in separate cryostats: the central one covers $|\eta|$ $\lsim 1.1$, and the two end 
sections extend the coverage to larger $\vert\eta\vert$. The calorimeter is segmented in depth, 
with four electromagnetic layers followed by up to five hadronic layers. It is  also segmented 
in projective towers of size $0.1\times 0.1$ in $\eta - \phi$ space, where $\phi$ is the azimuthal 
angle in radians. Calorimeter cells are defined as intersections of towers and layers. Additional 
sampling is provided by scintillating tiles in the regions at the boundary between cryostats.
An outer muon system, covering $|\eta|<2$, consists of a layer of tracking detectors and scintillation 
trigger counters in front of 1.8~T toroids, followed by two similar layers
after the toroids. Jets are reconstructed from the energy deposited in
calorimeter towers using the Run~II cone
algorithm~\cite{jetalgo} with radius $\mathcal{R}= \sqrt{(\Delta\phi)^2+(\Delta\eta)^2}=0.5$.
The jet energy scale (JES) is derived from the transverse momentum balance in photon-plus-jet 
events. The $\met$ is calculated from all calorimeter cells, and corrected for the jet energy 
scale and for reconstructed muons.

The D\O\ trigger system consists of three levels, L1, L2, and L3. 
The events used in this analysis were recorded using a jet trigger requiring missing transverse
energy calculated using the sum of the jet momenta \mbox{($\mht = \vert\sum_{\mathrm{jets}} \overrightarrow{p_T}\vert$).}
At L1, events were required to have at least three calorimeter towers of size 
$\Delta\phi \times \Delta\eta\,=\,0.2 \times 0.2$ with transverse energy $E_T$ greater than 5\,GeV.
Events with a large imbalance in transverse momentum were then selected by requiring $\mht$ to be greater than
20\,GeV and 30\,GeV at L2 and L3 respectively. In a small fraction of the data sample recorded at a higher 
instantaneous luminosity, the acoplanarity, defined as the azimuthal angle 
between the two leading jets, was required to be less than $168.75^\circ$ and $170^\circ$ at L2 and L3
respectively.

The signal consists of jets and $\met$. This topology also arises 
from SM processes with real $\met$, such as $p\bar p\to Z+$\,jets with $Z\to\nu\bar{\nu}$, and 
from multijet production when one or more jets are mismeasured (QCD background). Simulated 
events from SM and mSUGRA processes were produced using Monte Carlo (MC) 
generators, subjected to a full {\sc geant}-based~\cite{geant} simulation of the detector 
geometry and response, and processed through the same reconstruction chain as the data.
The {\sc CTEQ5L}~\cite{cteq5} 
parton density functions (PDF) were 
used, and a Poisson-average of 0.8 minimum bias events was overlaid on each simulated event.
The QCD background was not simulated, but estimated directly from data. 
To simulate $W/Z+$\,jets and $t\bar{t}$ production, the {\sc alpgen\,1.3} 
generator~\cite{Mangano:2002ea} was used, interfaced with {\sc pythia\,6.202}~\cite{Sjostrand:2000wi}
for the simulation of initial and final state radiation and of jet hadronization. 
The next-to-leading order (NLO) cross sections were computed with 
{\sc mcfm\,3.4.4}~\cite{Campbell:2001ik}, or taken from Ref.~\cite{Kidonakis:2004hr} for $t\bar{t}$ 
production.

Squark and gluino production and decay were simulated with {\sc pythia}.
The masses and couplings of the supersymmetric particles were calculated with 
{\sc isajet\,7.58} ~\cite{Paige:2003mg} from the set of five mSUGRA parameters: $m_0$ and $m_{1/2}$, 
which are universal scalar and gaugino masses, and $A_0$, a universal trilinear coupling, all 
defined at the scale of grand unification; $\tan\beta$, the ratio of the vacuum expectation 
values of the two Higgs fields; and the sign of the Higgs-mixing mass parameter $\mu$.
To retain consistency with earlier analyses~\cite{Abbott:1999xc,Affolder:2001tc}, 
the following parameters were fixed: $A_0 = 0$, $\tan\beta = 3$, and $\mu < 0$. For the same 
reason, the production of scalar top quarks, or stops, was ignored. In the following, 
``squark mass'' stands for the average mass of all squarks other than stops. All squark and 
gluino decay modes were taken into account in the simulation, including cascade decays such as
$\sg \to q\bar{q}\tilde\chi_2^0$ with $\tilde\chi_2^0\to\ell^+\ell^-\xo$.
The NLO cross sections of the various signal processes were calculated with 
{\sc prospino\,2}~\cite{Beenakker:1996ch}.

Three benchmark scenarios have been considered. At low $m_0$, the gluino is heavier than the 
squarks, and the process with the dominant cross section is $\sq\overline{\sq}$ production. 
A ``dijet'' analysis was optimized to search for events containing a pair of acoplanar jets.
At high $m_0$, the 
squarks are much heavier than the gluino, and the process with the highest cross section is 
therefore $\sg\sg$ production. A ``gluino'' analysis was optimized to search for multijet 
events ($\geq 4$ jets). In the intermediate $m_0$ region, all squark-gluino production processes
contribute to the total cross section, in particular the $\sq\sg$ process 
becomes relevant. A ``3-jets'' analysis was optimized to search for events with at least 
three jets. The benchmark for this analysis is the case where 
$m_{\sq}=m_{\sg}$.

A common event preselection was used for the three analyses to select events with at least 
two jets and substantial $\met$ ($\geq$ 40 GeV). 
The acoplanarity was required to be below $165^\circ$.
The longitudinal position of the primary vertex with respect to the detector center was 
restricted, $\vert z\vert < 60$\,cm, to ensure an efficient primary vertex reconstruction. 
The two leading jets, {\sl i.e.} those with the largest transverse energies,
were required to be in the central region of the calorimeter, $\etadet < 0.8$, where 
$\eta_{\mathrm{det}}$ is the jet pseudorapidity calculated under the assumption that the
jet originates from the detector center. These jets must have their
fraction of energy in the electromagnetic layers of the calorimeter smaller than 0.95.
Minimum transverse energies of 60 and 40\,GeV were required for the first and second leading 
jets, respectively.
 
The tracking capabilities of the Run~II D\O\ detector were used to significantly reduce the QCD 
background. 
A comparison of the jet energy with the energy carried by its associated charged 
particles was performed. 
In particular, the ratio {\sl CPF} of the transverse momentum carried by tracks associated with the jet
to the jet $E_T$ is expected to be close to zero if an incorrect primary 
vertex was selected. 
The two leading jets were required to have {\sl CPF} larger than 0.05.

\begin{table}
\caption{\label{cutflow1}
Selection criteria for the three analyses (all energies in GeV); see the text 
for further details.
}
\begin{ruledtabular}
\begin{tabular}{cccc}
Preselection Cut & \multicolumn{3}{c}{All Analyses} \\
\hline
$\met$				& \multicolumn{3}{c}{$\geq 40$}				\\
Acoplanarity			& \multicolumn{3}{c}{$< 165^\circ$}			\\
$|\mathrm{Vertex}\ z\ {\mathrm pos.}|$	& \multicolumn{3}{c}{$<60$ cm}				\\
\hline
Selection Cut			& ``dijet''	& ``3-jets''	& ``gluino'' 			\\
\hline
1st jet $E_T$\footnotemark[1]   & $\geq 60$ 	& $\geq 60$ 	& $\geq 60$		\\
2nd jet $E_T$\footnotemark[1]	& $\geq 50$ 	& $\geq 40$ 	& $\geq 40$		\\
3rd jet $E_T$\footnotemark[1]	& $-$		& $\geq 30$ 	& $\geq 30$		\\
4th jet $E_T$\footnotemark[1]	& $-$		& $-$		& $\geq 20$		\\
\hline
Electron veto			& yes		& yes		& yes			\\
Muon veto			& yes		& yes		& yes			\\
\hline
$\Delta\phi (\met,\mathrm{jet_1})$ 		& $\geq 90^\circ$ & $\geq 90^\circ$ & $\geq 90^\circ$ 	\\
$\Delta\phi (\met,\mathrm{jet_2})$ 		& $\geq 50^\circ$ & $\geq 50^\circ$ & $\geq 50^\circ$ 	\\
$\Delta\phi_{\rm{min}}(\met,\mathrm{any\,jet})$ & $\geq 40^\circ$	& $-$ & $-$				\\
\hline
$\HT$				& $\geq$ 275  & $\geq$ 350  & $\geq$ 225 		\\
$\met$			& $\geq$ 175  & $\geq$ 100  & $\geq$  75 		\\
\end{tabular}
\end{ruledtabular}
\footnotetext[1]{Jets subject to an $E_T$ cut are also required to be central \mbox{($\etadet<0.8$)}, 
with an electromagnetic fraction below 0.95, and to have {\sl CPF}$\geq$0.05.}
\end{table}

Different selection criteria were next applied in the three analyses, as summarized in 
Table~\ref{cutflow1}. In the ``dijet'' analysis, the cut on the second jet $E_T$ was 
raised to 50\,GeV. In the ``3-jets'' and ``gluino'' analyses, a third and fourth 
jet were required, respectively. They must fulfill the same quality criteria as the two 
leading jets, except for the $E_T$ cuts which were set at 30 and 20\,GeV. 
In all three analyses, a veto on isolated electrons or muons 
with $p_T>$10\,GeV rejects a large fraction of events originating from  the $W/Z+$\,jets 
processes. The azimuthal angles between the $\met$ and the first jet, 
$\Delta\phi(\met,\mathrm{jet_1})$, and the second jet, 
$\Delta\phi(\met,\mathrm{jet_2})$, were used to remove events where the energy of 
one jet was mismeasured, generating $\met$ aligned with that jet. The cuts are 
$\Delta\phi(\met,\mathrm{jet_1}) \geq 90^\circ$ and 
$\Delta\phi(\met,\mathrm{jet_2}) \geq 50^\circ$. 

\begin{figure*}
\begin{tabular}{ccc}
\includegraphics[width=5.5cm]{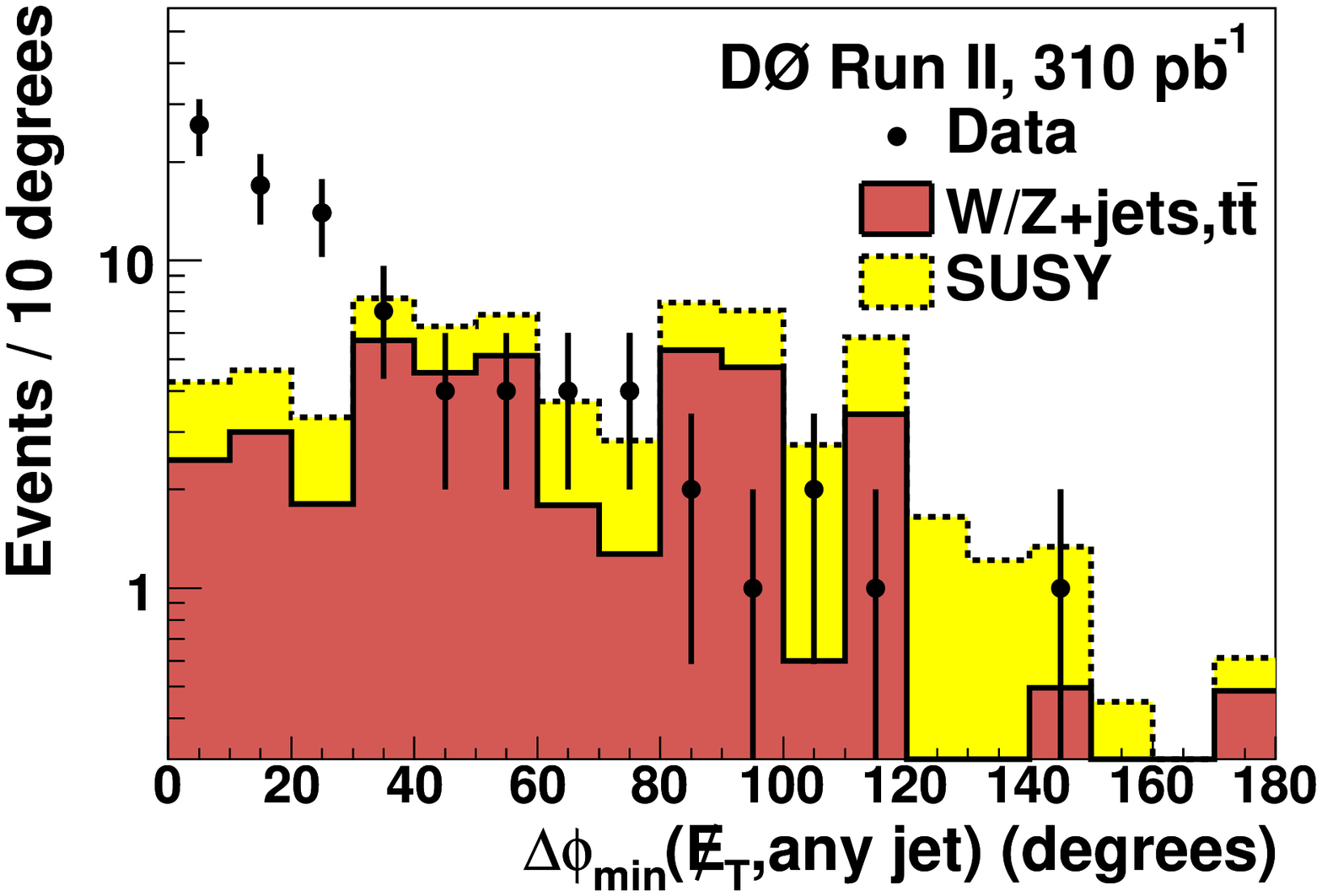} &
\includegraphics[width=5.5cm]{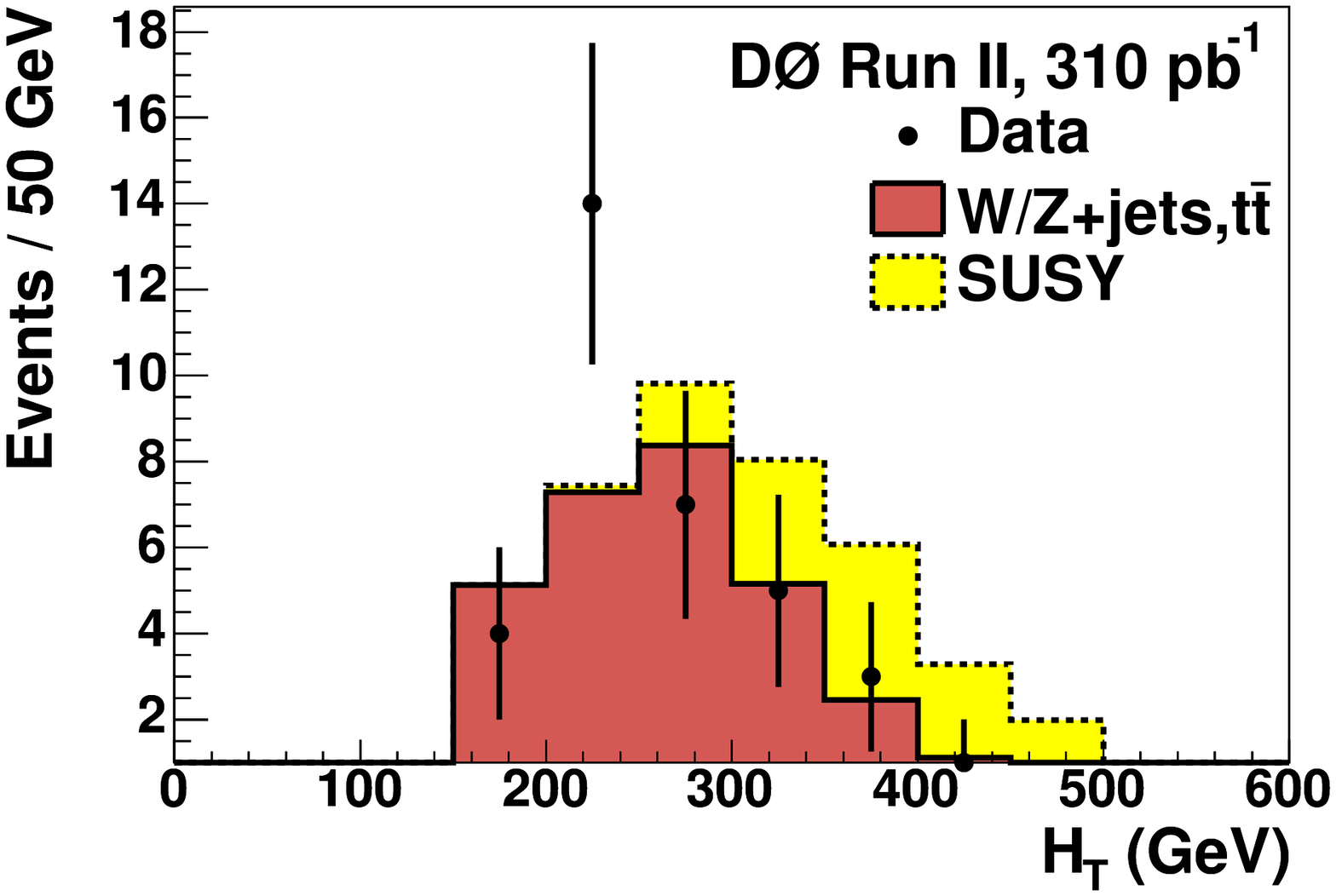}        &
\includegraphics[width=5.5cm]{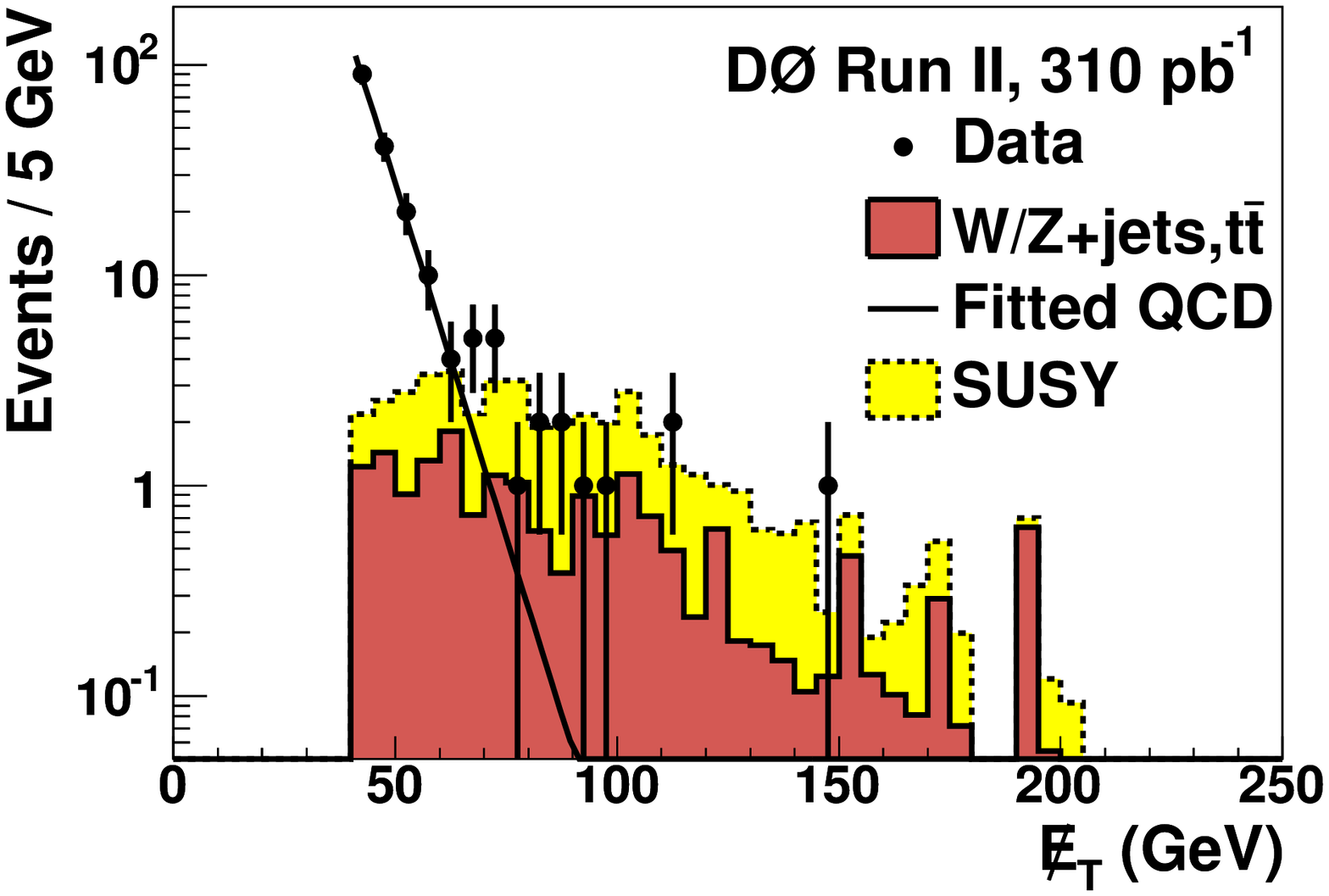} \\
\end{tabular}
\caption{\label{plots}
$\Delta\phi_{\rm{min}}(\met,\mathrm{any\,jet})$ distribution after applying the ``dijet'' analysis
criteria with a $\met$ cut reduced to 80\,GeV and without requiring the conditions on 
$\Delta\phi_{\rm{min}}(\met,\mathrm{any\,jet})$ itself and on $\Delta\phi(\met,\mathrm{jet_2})$ (left), 
$\HT$ distribution
after applying all the ``3-jets'' analysis criteria except the one on $\HT$ (middle), and $\met$ 
distribution after applying all the ``gluino'' analysis criteria except the final one on $\met$ (right),
for data (points with error bars), for non-QCD SM background (full histogram), 
and for signal MC (dashed histogram on top of SM).
For each analysis, the signal drawn is the one for the appropriate benchmark scenario (Table~\ref{summary}).
In the $\met$ distribution, the fitted QCD background is also drawn.
} 
\end{figure*}

\begin{table*}
\renewcommand{\arraystretch}{1.2}
\begin{center}
\caption{\label{summary}
For each analysis, information on the signal for which it was optimized: 
$m_0$, $m_{1/2}$, $m_{\sg}$, $m_{\sq}$ and nominal NLO cross section,
signal efficiency, the number of events observed, the number of events 
expected from SM and QCD backgrounds and the 95\% C.L. signal cross section upper limit.
The first uncertainty is statistical and the second is systematic.
}
\begin{ruledtabular}
\begin{tabular}{lccccccc}
Analysis   & $(m_0,m_{1/2})$ & $(m_{\sg},m_{\sq})$ & $\sigma_{\mathrm{nom}}$ & $\epsilon_{\mathrm{sig.}}$     & $N_{\mathrm{obs.}}$ & $N_{\mathrm{backgrd.}}$ 		& $\sigma_{95}$\\
           & (GeV)           & (GeV)               & (pb)                   & (\%)                             &            	    &                         		& (pb) \\
\hline
``dijet''  & (25,145)        & (366,318)           & 0.63  	             & $6.2 \pm 0.4 ^{+1.1}_{-0.9}$ &  6             & $ 4.8^{+4.4}_{-2.0}\ ^{+1.1}_{-0.8}$	& 0.44 \\
``3-jets'' & (191,126)       & (330,330)           & 0.64  	             & $4.7 \pm 0.3 ^{+0.8}_{-0.7}$ &  4             & $ 3.9^{+1.3}_{-1.0}\ ^{+0.7}_{-0.8}$	& 0.45 \\
``gluino'' & (500,80)        & (240,507)           & 2.41  	             & $2.3 \pm 0.2 ^{+0.4}_{-0.3}$ & 10             & $10.3^{+1.5}_{-1.4}\ ^{+1.9}_{-2.5}$	& 1.72 \\
\end{tabular}
\end{ruledtabular}
\end{center}
\end{table*}

In the ``dijet'' analysis, QCD events were further suppressed by requiring that the minimum azimuthal angle
$\Delta\phi_{\rm{min}}(\met,\mathrm{any\,jet})$ between the $\met$ and any jet
with $E_T>15$\,GeV be greater than $40^{\circ}$. Because of the higher jet multiplicity, 
this criterion was not used in the ``3-jets'' and ``gluino'' analyses.

The ``dijet'' $\Delta\phi_{\rm{min}}(\met,\mathrm{any\,jet})$ cut along with the two final cuts 
on $\HT = \sum_{\mathrm{jets}} E_T$ and on $\met$ were optimized by minimizing 
the expected upper limit on the cross section in the absence of signal.
To this end, 
as well as for the derivation of the final results, the modified frequentist {\sl CL}$_s$ method~\cite{CLS}
was used. For each set of cuts tested, the QCD background contribution was estimated from 
an exponential fit to the $\met$ distribution below 60\,GeV, after subtraction of the SM 
background processes, extrapolated above the chosen $\met$ cut value. The optimal cuts thus 
determined are given in Table~\ref{cutflow1} for the three analyses. 
Figure~\ref{plots} shows: the $\Delta\phi_{\rm{min}}(\met,\mathrm{any\,jet})$ distribution after
applying the ``dijet'' analysis criteria with a $\met$ cut reduced to 80\,GeV and without requiring 
the conditions on $\Delta\phi_{\rm{min}}(\met,\mathrm{any\,jet})$ itself and on
$\Delta\phi(\met,\mathrm{jet_2})$; the $\HT$ distribution after applying all the ``3-jets'' analysis 
criteria except the one on $\HT$; and the \,$\met$ distribution after applying all the ``gluino'' 
analysis criteria except the one on $\met$.

The numbers of events selected by each analysis are reported in Table~\ref{summary}, 
as well as the numbers of background events expected.
Six events were selected by the ``dijet'' analysis, four by the ``3-jets'' analysis, and ten 
by the ``gluino'' analysis. The total expected 
background contributions are 4.8, 3.9 and 10.3 events, respectively. The main background 
contributions are from $\Z\to\nu\bar{\nu}+\mathrm{jets}$, $\W\to l\nu+\mathrm{jets}$, and
\ttb\,$\to$\,\bbb\,$q\bar{q}'l\nu$.
The QCD background was evaluated from a fit to the $\met$ distribution as described above.
It was found to be negligible in the ``dijet'' and ``3-jets'' analyses, 
and was therefore conservatively ignored. A QCD contribution of $0.7^{+0.7}_{-0.4}$ event was estimated 
in the ``gluino'' analysis. The uncertainties were obtained by taking into account the accuracy of the
fit parameter determination and by varying the range of the fit.
The signal efficiencies are given in Table~\ref{summary} 
for the three benchmark scenarios, with the corresponding values of $m_0$, $m_{1/2}$,
the squark and gluino masses, and the NLO cross section. 
The quoted systematic uncertainties are discussed below.

The uncertainty coming from the JES corrections is one of the most important. 
It is typically of the order of 13\% for the SM backgrounds and
10\% for the signal efficiencies.
The uncertainties on the jet energy resolution, on the jet track confirmation, and on the jet 
reconstruction and identification efficiencies were evaluated. They lead to systematic 
uncertainties of 3.5\%, 4.0\% and 5.4\% in the ``dijet,'' ``3-jets,'' and ``gluino'' analyses, respectively. 
The trigger was found to be fully efficient for the event samples surviving all analysis cuts.
Conservatively, a 2\% uncertainty was set on the trigger efficiency.
The uncertainty on the determination of the luminosity is 6.5\%~\cite{d0lumi}.
All of these uncertainties are fully correlated between signal and SM backgrounds. 
A 15\% systematic uncertainty was set on the $W/Z$+jets and \ttb\ NLO cross sections.
The uncertainty on the signal acceptance due to the PDF choice was 
determined to be 6\%, using the forty-eigenvector basis of the {\sc CTEQ6.1M} PDF set~\cite{Pumplin:2002vw}.

\begin{figure*}
\begin{tabular}{ccc}
\includegraphics[width=5.5cm]{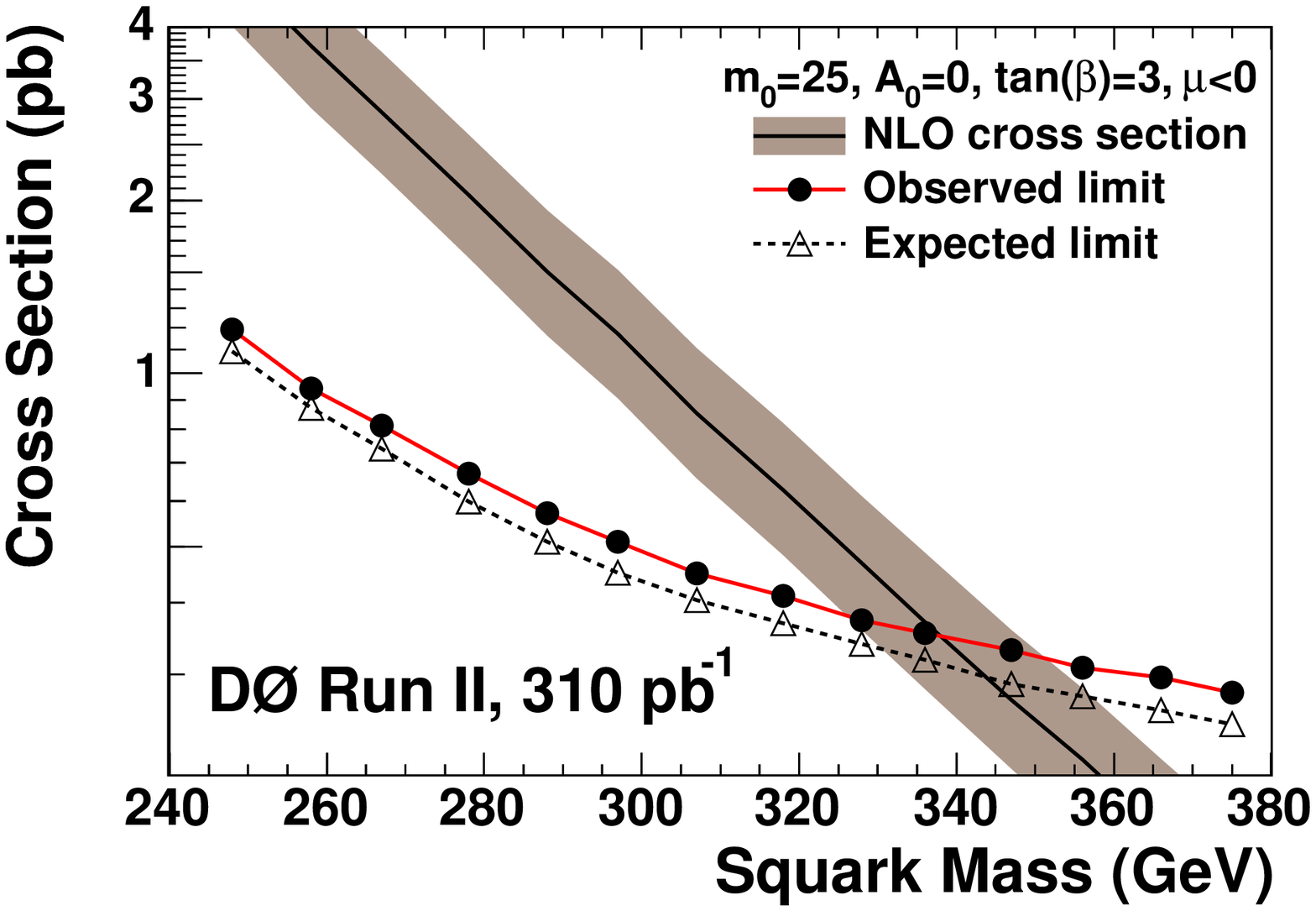} &
\includegraphics[width=5.5cm]{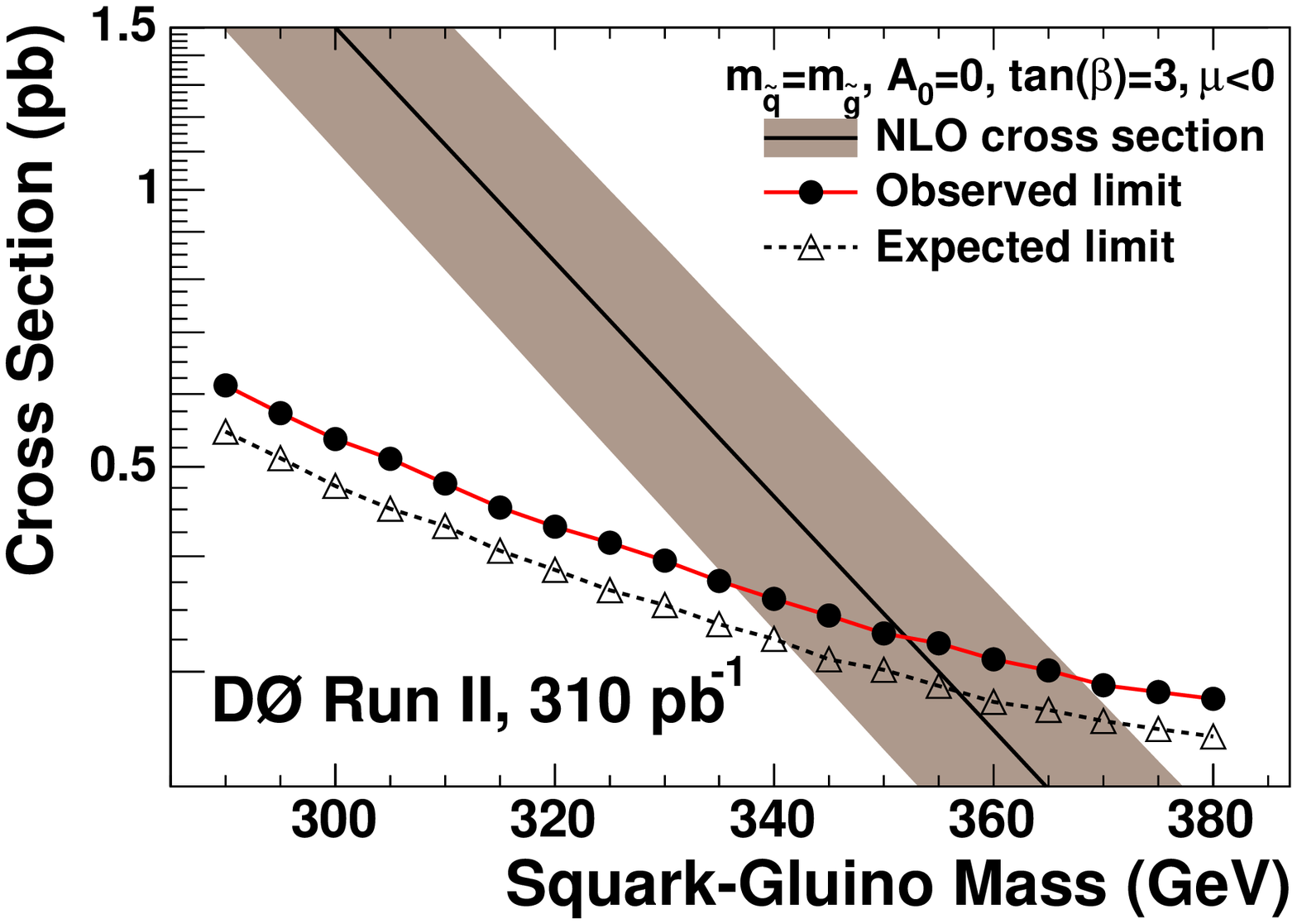} &
\includegraphics[width=5.5cm]{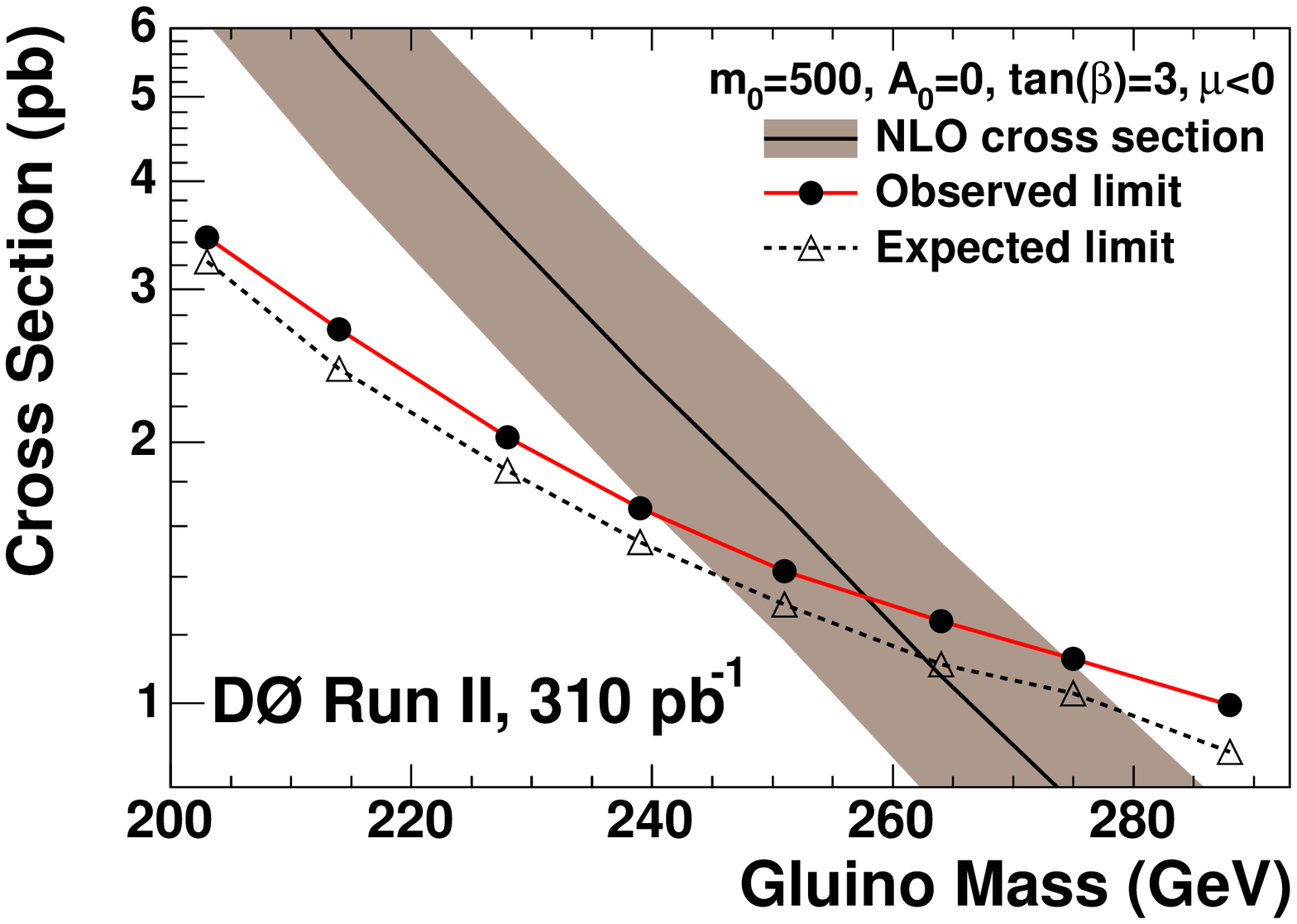} \\
\end{tabular}
\caption{\label{xseclim}
For $\tan\beta = 3$, $A_0 = 0$, $\mu < 0$, observed (closed circles) and expected (opened triangles) 
95\% C.L. upper limits on squark-gluino production cross sections combining the analyses for 
$m_0=25$\,GeV (left), $m_{\sq}=m_{\sg}$ (middle), and $m_0=500$\,GeV (right). The nominal production 
cross sections are also shown, with shaded bands corresponding to the PDF and 
renormalization-and-factorization scale uncertainties.
} 
\end{figure*}

The signal cross sections are very sensitive to the PDF choice and to the renormalization 
and factorization scale, $\mu_{\mathrm{rf}}$.
The nominal NLO cross sections, $\sigma_{\mathrm{nom}}$, were computed with the 
{\sc CTEQ6.1M} PDF and for $\mu_{\mathrm{rf}}=Q$, where $Q$ was taken to be equal 
to $m_{\sg}$ for $\sg\sg$ production, $m_{\sq}$ for $\sq\sq$ and $\sq\overline{\sq}$ productions, 
and $(m_{\sq}+m_{\sg})/2$ for $\sq\sg$ production. 
The uncertainty due to the choice of PDF was determined using the full set
of {\sc CTEQ6.1M} eigenvectors, with the individual uncertainties added in
quadrature.
The effect on the nominal signal cross 
sections, which varies between 15\% and 50\%, is dominated by the large uncertainty on the gluon 
distribution at \mbox{high $x$}. The effect of the renormalization and factorization scale was studied 
by calculating the signal cross sections for $\mu_{\mathrm{rf}}=Q$, $\mu_{\mathrm{rf}}=Q/2$ and 
$\mu_{\mathrm{rf}}=2 \times Q$.
The factor two on this scale reduces or increases the nominal signal cross sections by \mbox{15$-$20\%}.
The PDF and $\mu_{\mathrm{rf}}$ effects were added in quadrature to compute minimum, 
$\sigma_{\mathrm{min}}$, and maximum, $\sigma_{\mathrm{max}}$, signal cross sections. 

No significant excess of events was observed in the data with respect to the SM background 
expectation in any of the three analyses. Therefore, an excluded domain in the gluino-squark
mass plane was determined as follows.
The three analyses were run over signal MC samples generated in the gluino-squark mass plane 
to compute signal efficiencies. Then, to take advantage of the different features of the three 
analyses, they were combined in the limit computation, with the small overlaps taken  into account.
In the data, no events were selected by more than one analysis.

Limits at the 95\% C.L. were computed for three hypotheses on the signal cross sections:
nominal, minimum, and maximum. Figure~\ref{xseclim} shows the observed and expected 
upper limits on squark-gluino production cross sections for the three benchmark 
scenarios.
For the ``3-jets'' and ``gluino'' analyses, the expected limits computed 
with the numbers of events reported in Table~\ref{summary} are almost 
identical to the observed ones. Once the combination of analyses is 
performed, the expected limits become slightly better than the observed 
limits at large $m_0$ and for $m_{\sq}=m_{\sg}$, as can be seen in 
Figure~\ref{xseclim}.

\begin{figure}
\includegraphics[width=8.5cm]{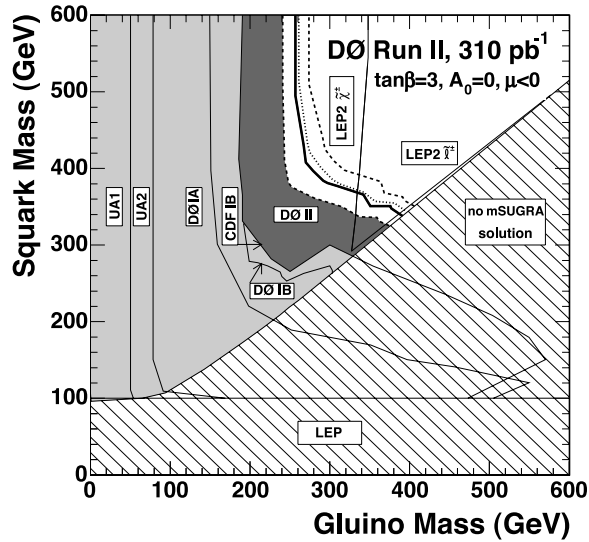}
\caption{\label{contour}
In the gluino and squark mass plane, excluded regions at the 95\% C.L.
by direct searches in the mSUGRA framework with
$\tan\beta = 3$, $A_0 = 0$, $\mu < 0$. 
The new region excluded by this analysis in the most conservative 
hypothesis ($\sigma_{\mathrm{min}}$) is shown in dark shading.
The thick line is the limit of the excluded region for the $\sigma_{\mathrm{nom}}$ hypothesis.
The corresponding expected limit is the dotted line.
The band delimited by the two dashed lines shows the effect of the PDF choice and of a variation of 
$\mu_{\mathrm{rf}}$ by a factor of two. 
Regions excluded by previous experiments are indicated in light 
shading~\cite{Abbott:1999xc,Affolder:2001tc,prevexp}. 
The two thin lines indicate the indirect limits inferred from the LEP2 chargino
and slepton searches~\cite{lepindirect}.
The region where no mSUGRA solution can be found is shown hatched.
}
\end{figure}

Figure~\ref{contour} shows the excluded domain in the gluino-squark mass plane.
The absolute lower limits on the squark and gluino masses obtained in the most conservative 
hypothesis, $\sigma_{\mathrm{min}}$, are 325~GeV and 241~GeV, respectively. The corresponding 
expected limits are 330\,GeV and 246\,GeV. Table~\ref{finallimits2} summarizes these absolute 
limits as a function of the signal cross section hypothesis.
Limits were also derived for the particular case $m_{\sq}=m_{\sg}$. For 
$\sigma_{\mathrm{min}}$, squark and gluino masses below 337\,GeV are excluded, while 
the expected limit is 340\,GeV. The observed limit becomes 351\,GeV for 
$\sigma_{\mathrm{nom}}$, and 368\,GeV for $\sigma_{\mathrm{max}}$.

These results improve on the previous direct limits on squark and gluino 
masses~\cite{Abbott:1999xc,Affolder:2001tc,prevexp}.
They were obtained within the mSUGRA framework with $\tan\beta = 3$, $A_0 = 0$, and
$\mu < 0$. 
A general scan of the mSUGRA parameter space is beyond the scope of the current analysis, but 
it has been verified that similar results would be obtained for a large class of parameter 
sets. 
The limits obtained at LEP on the chargino ($\xpm$) and slepton ($\tilde{\ell}$) masses can be 
turned into constraints on the mSUGRA parameters $m_0$ and $m_{1/2}$~\cite{lepindirect}, and
hence on the squark and gluino masses as shown in Fig.~\ref{contour}. 
The limits from Higgs boson searches at LEP are even more constraining~\cite{lepindirect}, 
actually ruling out all of the squark and gluino mass domain to which the Tevatron could be 
sensitive. The interpretation of these indirect
constraints is however more sensitive to the details of the model considered than the direct 
limits presented here.

\begin{table}
\renewcommand{\arraystretch}{1.2}
\begin{center}
\caption{\label{finallimits2}
Absolute lower limits at the 95\% C.L. on the squark and gluino masses (in GeV) as a function 
of the choice of signal cross section hypothesis as defined in the text. 
Numbers in parentheses correspond to the expected limits. These limits are valid for the 
mSUGRA parameters: $\tan\beta = 3$, $A_0 = 0$, $\mu < 0$.
}
\begin{ruledtabular}
\begin{tabular}{lcc}
Hypothesis	   		&  Gluino mass 	& Squark mass \\
\hline
$\sigma_{\mathrm{min}}$ 	& 241   (246)	& 325 (330)  \\
$\sigma_{\mathrm{nom}}$ 	& 257   (261)	& 339 (344)  \\
$\sigma_{\mathrm{max}}$   	& 274   (280)	& 352 (358)  \\
\end{tabular}
\end{ruledtabular}
\end{center}
\end{table}

In summary, a search for events with jets and large $\met$ has been
performed in a 310~$\mathrm{pb}^{-1}$ data sample from $p\bar{p}$
collisions at 1.96\,TeV, collected by the D\O\ detector. Three analyses
were designed, specifically targeted to the dijet, three-jet, and
multijet topologies. The numbers of events observed are in agreement
with the SM background predictions. The results have been interpreted in
the framework of minimal supergravity with $\tan\beta = 3$, $A_0 = 0$,
$\mu < 0$. For the central choice of PDF, and for a renormalization and
factorization scale equal to the mass of the squark or gluino produced,
the lower limits on the squark and gluino masses are 339 and 257~GeV at
the 95\% C.L. Taking into account the PDF uncertainties and allowing for
a factor of two in the choice of scale, these limits are reduced to 325
and 241 GeV, respectively. These are the most constraining direct limits 
on the squark and gluino masses to date.

\input acknowledgement_paragraph_r2.tex   

\end{document}

%% file: list_of_authors_r2.tex
%
\author{                                                                      
V.M.~Abazov,$^{36}$                                                           
B.~Abbott,$^{76}$                                                             
M.~Abolins,$^{66}$                                                            
B.S.~Acharya,$^{29}$                                                          
M.~Adams,$^{52}$                                                              
T.~Adams,$^{50}$                                                              
M.~Agelou,$^{18}$                                                             
J.-L.~Agram,$^{19}$                                                           
S.H.~Ahn,$^{31}$                                                              
M.~Ahsan,$^{60}$
% ...
\\                                                                            
\vskip 0.30cm                                                                 
\centerline{(D\O\ Collaboration)}                                             
\vskip 0.30cm                                                                 
}                                                                             
\affiliation{                                                                 
\centerline{$^{1}$Universidad de Buenos Aires, Buenos Aires, Argentina}       
\centerline{$^{2}$LAFEX, Centro Brasileiro de Pesquisas F{\'\i}sicas,         
                  Rio de Janeiro, Brazil}                                     
\centerline{$^{3}$Universidade do Estado do Rio de Janeiro,                   
                  Rio de Janeiro, Brazil}                                     
\centerline{$^{4}$Instituto de F\'{\i}sica Te\'orica, Universidade            
                  Estadual Paulista, S\~ao Paulo, Brazil}                     
\centerline{$^{5}$University of Alberta, Edmonton, Alberta, Canada,           
                  Simon Fraser University, Burnaby, British Columbia, Canada,}
\centerline{York University, Toronto, Ontario, Canada, and                    
                  McGill University, Montreal, Quebec, Canada}                
\centerline{$^{6}$Institute of High Energy Physics, Beijing,                  
                  People's Republic of China}                                 
\centerline{$^{7}$University of Science and Technology of China, Hefei,       
                  People's Republic of China}                                 
\centerline{$^{8}$Universidad de los Andes, Bogot\'{a}, Colombia}             
\centerline{$^{9}$Center for Particle Physics, Charles University,            
                  Prague, Czech Republic}                                     
\centerline{$^{10}$Czech Technical University, Prague, Czech Republic}        
\centerline{$^{11}$Center for Particle Physics, Institute of Physics,         
                   Academy of Sciences of the Czech Republic,                 
                   Prague, Czech Republic}                                    
\centerline{$^{12}$Universidad San Francisco de Quito, Quito, Ecuador}        
\centerline{$^{13}$Laboratoire de Physique Corpusculaire, IN2P3-CNRS,         
                   Universit\'e Blaise Pascal, Clermont-Ferrand, France}      
\centerline{$^{14}$Laboratoire de Physique Subatomique et de Cosmologie,      
                   IN2P3-CNRS, Universite de Grenoble 1, Grenoble, France}    
\centerline{$^{15}$CPPM, IN2P3-CNRS, Universit\'e de la M\'editerran\'ee,     
                   Marseille, France}                                         
\centerline{$^{16}$IN2P3-CNRS, Laboratoire de l'Acc\'el\'erateur              
                   Lin\'eaire, Orsay, France}                                 
\centerline{$^{17}$LPNHE, IN2P3-CNRS, Universit\'es Paris VI and VII,         
                   Paris, France}                                             
\centerline{$^{18}$DAPNIA/Service de Physique des Particules, CEA, Saclay,    
                   France}                                                    
\centerline{$^{19}$IReS, IN2P3-CNRS, Universit\'e Louis Pasteur, Strasbourg,  
                    France, and Universit\'e de Haute Alsace,                 
                    Mulhouse, France}                                         
\centerline{$^{20}$Institut de Physique Nucl\'eaire de Lyon, IN2P3-CNRS,      
                   Universit\'e Claude Bernard, Villeurbanne, France}         
\centerline{$^{21}$III. Physikalisches Institut A, RWTH Aachen,               
                   Aachen, Germany}                                           
\centerline{$^{22}$Physikalisches Institut, Universit{\"a}t Bonn,             
                   Bonn, Germany}                                             
\centerline{$^{23}$Physikalisches Institut, Universit{\"a}t Freiburg,         
                   Freiburg, Germany}                                         
\centerline{$^{24}$Institut f{\"u}r Physik, Universit{\"a}t Mainz,            
                   Mainz, Germany}                                            
\centerline{$^{25}$Ludwig-Maximilians-Universit{\"a}t M{\"u}nchen,            
                   M{\"u}nchen, Germany}                                      
\centerline{$^{26}$Fachbereich Physik, University of Wuppertal,               
                   Wuppertal, Germany}                                        
\centerline{$^{27}$Panjab University, Chandigarh, India}                      
\centerline{$^{28}$Delhi University, Delhi, India}                            
\centerline{$^{29}$Tata Institute of Fundamental Research, Mumbai, India}     
\centerline{$^{30}$University College Dublin, Dublin, Ireland}                
\centerline{$^{31}$Korea Detector Laboratory, Korea University,               
                   Seoul, Korea}                                              
\centerline{$^{32}$SungKyunKwan University, Suwon, Korea}                     
\centerline{$^{33}$CINVESTAV, Mexico City, Mexico}                            
\centerline{$^{34}$FOM-Institute NIKHEF and University of                     
                   Amsterdam/NIKHEF, Amsterdam, The Netherlands}              
\centerline{$^{35}$Radboud University Nijmegen/NIKHEF, Nijmegen, The          
                  Netherlands}                                                
\centerline{$^{36}$Joint Institute for Nuclear Research, Dubna, Russia}       
\centerline{$^{37}$Institute for Theoretical and Experimental Physics,        
                   Moscow, Russia}                                            
\centerline{$^{38}$Moscow State University, Moscow, Russia}                   
\centerline{$^{39}$Institute for High Energy Physics, Protvino, Russia}       
\centerline{$^{40}$Petersburg Nuclear Physics Institute,                      
                   St. Petersburg, Russia}                                    
\centerline{$^{41}$Lund University, Lund, Sweden, Royal Institute of          
                   Technology and Stockholm University, Stockholm,            
                   Sweden, and}                                               
\centerline{Uppsala University, Uppsala, Sweden}                              
\centerline{$^{42}$Physik Institut der Universit{\"a}t Z{\"u}rich,            
                   Z{\"u}rich, Switzerland}                                   
\centerline{$^{43}$Lancaster University, Lancaster, United Kingdom}           
\centerline{$^{44}$Imperial College, London, United Kingdom}                  
\centerline{$^{45}$University of Manchester, Manchester, United Kingdom}      
\centerline{$^{46}$University of Arizona, Tucson, Arizona 85721, USA}         
\centerline{$^{47}$Lawrence Berkeley National Laboratory and University of    
                   California, Berkeley, California 94720, USA}               
\centerline{$^{48}$California State University, Fresno, California 93740, USA}
\centerline{$^{49}$University of California, Riverside, California 92521, USA}
\centerline{$^{50}$Florida State University, Tallahassee, Florida 32306, USA} 
\centerline{$^{51}$Fermi National Accelerator Laboratory,                     
            Batavia, Illinois 60510, USA}                                     
\centerline{$^{52}$University of Illinois at Chicago,                         
            Chicago, Illinois 60607, USA}                                     
\centerline{$^{53}$Northern Illinois University, DeKalb, Illinois 60115, USA} 
\centerline{$^{54}$Northwestern University, Evanston, Illinois 60208, USA}    
\centerline{$^{55}$Indiana University, Bloomington, Indiana 47405, USA}       
\centerline{$^{56}$University of Notre Dame, Notre Dame, Indiana 46556, USA}  
\centerline{$^{57}$Purdue University Calumet, Hammond, Indiana 46323, USA}    
\centerline{$^{58}$Iowa State University, Ames, Iowa 50011, USA}              
\centerline{$^{59}$University of Kansas, Lawrence, Kansas 66045, USA}         
\centerline{$^{60}$Kansas State University, Manhattan, Kansas 66506, USA}     
\centerline{$^{61}$Louisiana Tech University, Ruston, Louisiana 71272, USA}   
\centerline{$^{62}$University of Maryland, College Park, Maryland 20742, USA} 
\centerline{$^{63}$Boston University, Boston, Massachusetts 02215, USA}       
\centerline{$^{64}$Northeastern University, Boston, Massachusetts 02115, USA} 
\centerline{$^{65}$University of Michigan, Ann Arbor, Michigan 48109, USA}    
\centerline{$^{66}$Michigan State University,                                 
            East Lansing, Michigan 48824, USA}                                
\centerline{$^{67}$University of Mississippi,                                 
            University, Mississippi 38677, USA}                               
\centerline{$^{68}$University of Nebraska, Lincoln, Nebraska 68588, USA}      
\centerline{$^{69}$Princeton University, Princeton, New Jersey 08544, USA}    
\centerline{$^{70}$State University of New York, Buffalo, New York 14260, USA}
\centerline{$^{71}$Columbia University, New York, New York 10027, USA}        
\centerline{$^{72}$University of Rochester, Rochester, New York 14627, USA}   
\centerline{$^{73}$State University of New York,                              
            Stony Brook, New York 11794, USA}                                 
\centerline{$^{74}$Brookhaven National Laboratory, Upton, New York 11973, USA}
\centerline{$^{75}$Langston University, Langston, Oklahoma 73050, USA}        
\centerline{$^{76}$University of Oklahoma, Norman, Oklahoma 73019, USA}       
\centerline{$^{77}$Oklahoma State University, Stillwater, Oklahoma 74078, USA}
\centerline{$^{78}$Brown University, Providence, Rhode Island 02912, USA}     
\centerline{$^{79}$University of Texas, Arlington, Texas 76019, USA}          
\centerline{$^{80}$Southern Methodist University, Dallas, Texas 75275, USA}   
\centerline{$^{81}$Rice University, Houston, Texas 77005, USA}                
\centerline{$^{82}$University of Virginia, Charlottesville,                   
            Virginia 22901, USA}                                              
\centerline{$^{83}$University of Washington, Seattle, Washington 98195, USA}  
}                                                                             

%% file: acknowledgement_paragraph_r2.tex
%
We thank the staffs at Fermilab and collaborating institutions, 
and acknowledge support from the 
DOE and NSF (USA);
CEA and CNRS/IN2P3 (France);
FASI, Rosatom and RFBR (Russia);
CAPES, CNPq, FAPERJ, FAPESP and FUNDUNESP (Brazil);
DAE and DST (India);
Colciencias (Colombia);
CONACyT (Mexico);
KRF and KOSEF (Korea);
CONICET and UBACyT (Argentina);
FOM (The Netherlands);
PPARC (United Kingdom);
MSMT (Czech Republic);
CRC Program, CFI, NSERC and WestGrid Project (Canada);
BMBF and DFG (Germany);
SFI (Ireland);
The Swedish Research Council (Sweden);
Research Corporation;
Alexander von Humboldt Foundation;
and the Marie Curie Program.